# Comparison Study of Inertial Sensor Signal Combination for Human Activity Recognition based on Convolutional Neural Networks

Farhad Nazari, *Member,* IEEE[*], Navid Mohajer, Darius Nahavandi, *Member,* IEEE, Abbas Khosravi, *Senior Member,* IEEE, and Saeid Nahavandi, *Fellow,* IEEE

Institute for Intelligent System Research and Innovation (IISRI), Deakin University, Australia

*Abstract*— Human Activity Recognition (HAR) is one of the essential building blocks of so many applications like security, monitoring, the internet of things and human-robot interaction. The research community has developed various methodologies to detect human activity based on various input types. However, most of the research in the field has been focused on applications other than human-in-the-centre applications. This paper focused on optimising the input signals to maximise the HAR performance from wearable sensors. A model based on Convolutional Neural Networks (CNN) has been proposed and trained on different signal combinations of three Inertial Measurement Units (IMU) that exhibit the movements of the dominant hand, leg and chest of the subject. The results demonstrate k-fold cross-validation accuracy between 99.77 and 99.98% for signals with the modality of 12 or higher. The performance of lower dimension signals, except signals containing information from both chest and ankle, was far inferior, showing between 73 and 85% accuracy.

## I. Introduction

Human Activity Recognition (HAR) is a special application of pattern recognition that focuses on classifying human motions [1]. Various applications rely on this technology to perform well. Some of these applications are gaming [2], internet of things and smart homes [3], health care [4] and elderly care, human-robot interaction [5], monitoring and security [6] and other data mining applications [7].

Rawashdeh et al. proposed a HAR model for smart homes based on activity profiling [8]. Zhang et al. developed a Deep Learning (DL) framework for HAR based on smart sensors [9]. They utilised a hybrid Convolutional Neural Network (CNN); a dynamic platform (D-CNN), and a sensor (S-CNN), to improve the performance of the classifier. Chapron et al. built a wearable prototype to recognise daily activity and exercise in real-time from an Inertial Measurement Unit (IMU) [10]. Bonnechere et al. proposed a skeleton detection model for the human upper-limb motion from Microsoft Kinnect to evaluate rehabilitation exercises [11].

Different techniques can help with the detection of human activity from the input signal. Image processing is a widely used method for unusual activity detection [12]. Signal processing is another approach that uses time-series sensor signals like the ones embedded in smartphones [13] and watches [14]. Another widely used input is surface electromyogram (sEMG) [15]. Tuncer et al. proposed a model to classify hand movement to help control prosthetic hands [16]. They used ternary pattern and discrete wavelet to extract features from EMG signals and achieved up to 99.1% accuracy. Gu et al. proposed a new model-based stacked denoising autoencoder for locomotion activity recognition [17]. They used four different types of sensors and showed that one could achieve better results using a combination of sensors.

Different algorithms and methodologies can be used to detect the activity from the input data. Vrigkas et al. categorised HAR into unimodal and multi-modal activity recognition [18]. They further divided unimodal approaches into space-time [19], stochastic [20], rule-based [21], and shape-based [22] methods. Multi-modal HAR uses data from various sources of data with different or the same modality. In the past two decades, Machine Learning (ML) techniques like Gradient Boosting (GB) [23], Linear Discriminant Analysis (LDA) [24], K-Nearest Neighbors (KNN) [25], DT [26] and Support Vector Machine (SVM) [27] have gotten popularity in the field of HAR. However, they rely on mathematical techniques to extract features for each input type and lack a systematic approach [28].

Deep learning techniques aim to solve this problem by extracting high-level features from the raw or pre-processed input data. Oniga and Suto optimised the sensor configuration and developed a two-layer perceptron Artificial Neural Network (ANN) for implementation on FPGA [29]. Inoue et al. achieved up to 71% better recognition rate with a Deep Recurrent Neural Network (DRNN) than traditional ML models [30]. Cho and Yoon used a hierarchical Convolutional Neural Network (CNN) to increase the prediction accuracy [31]. Xia et al. proposed a model consisting of Long Short-Term Memory (LSTM) layers followed by convolution layers, reaching up to 95.8% $F_1$ score [32]. Combining CNN and Gated Recurrent Unit (GRU) helped Dua et al. to achieve maximum of 97.2% classification accuracy [33].

HAR also play an important role in wearable assistive devices, e.g. orthoses and exoskeletons. Knowing the user's activity and state can help adjust the assistance level of a semi-passive device or improve the real-time control of an active one [34]. Poliero et al. attempted to enhance the versatility of a back-support exoskeleton using HAR [35]. Zheng et al. tried to detect the motion of the human's torso from EMG signals and a Decision Tree (DT) classifier [36]. However, despite the size of the literature review in HAR, not enough research has been conducted to accommodate the activity detection for the application of wearable assistive devices. That is why the developed knowledge may not be easy to implement in the control system of these devices. This restriction shows itself in the sensor type and location, duration and methodology of feature extraction, and predefined activities.

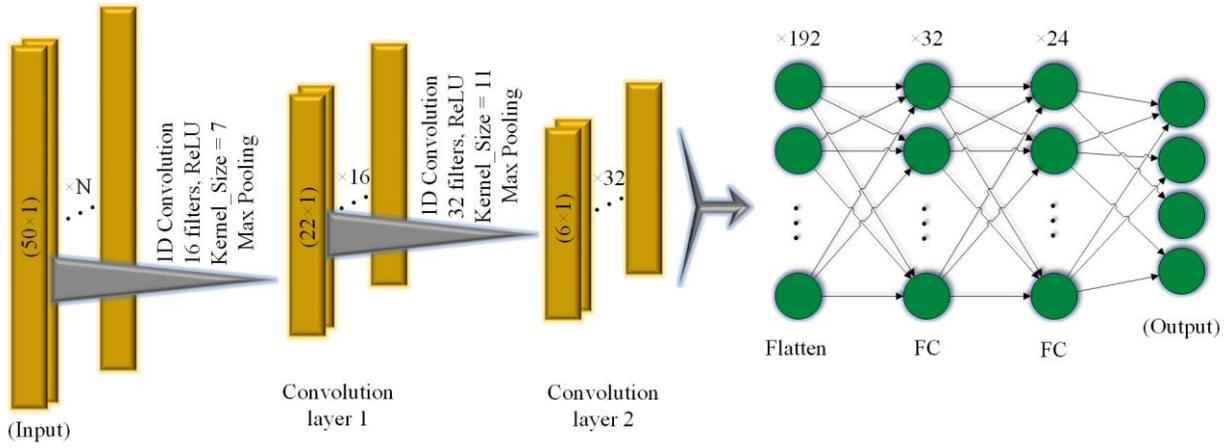

Fig. 1. The structure of the proposed DCNN model. N is the modality of the input data which is three for each sensor used in each trial. For example in the gyro data of all three sensors, the modality is nine.

In this study, we aim to optimise the input inertial data to detect the human activity for the application of a wearable assistive device, e.g. prosthesis or exoskeleton. The targets are signal type, modality and sensor location. A CNN model has been developed to detect a user's routine daily activity from the inertial motion data. The advantage of IMUs compared to other widely used sensors in the field like EMG is that the sensor does not require to be attached to the user's skin and can be installed in the device parts or the user's clothes.

The next section describes the dataset and data selection, and combination methodology. In section III, the proposed model will be described. Then the results will be discussed in section VI. And section V concludes this research.

## II. DATASET AND DATA PROCESSING

This section describes the dataset selection and processing methods used in this study.

### A. Dataset and Data Selection

In this study, we used the PAMAP2 Physical Activity Monitoring dataset has been used in [37]. This dataset has nine participants performing 18 activities. Subjects were wearing three IMU sensors on the wrist, chest and ankle and a heart rate monitor during the data collection experiment. The IMUs record the temperature, acceleration, gyroscope and magnetometer data in three dimensions at 100Hz. The heart-rate sensor records data with a 9Hz frequency, which was later up-sampled to match the IMU data. The hand and ankle inertial signals have been collected from the dominant limbs of the participants.

In this study, we used the gyrometer and accelerometer signals. The temperature doesn't seem to vary in a meaningful way during activities. Also, due to the need for regular calibration of the magnetometer, we decided that this type of sensor might not be the best candidate for our application. Also, for the purpose of user comfort and not attaching any sensor directly to their body, we did not use the physiological data.

### B. Data Preparation

A flaw of the dataset collected by Reiss and Stricker [38] is that the subjects' participation time in different activities is not uniform. Also, not all activities have a meaningful difference in the control of the assistive device. For these reasons, we selected five daily activities: sitting, standing, walking, ascending and descending stairs.

The following combinations of signals have been pre-processed and used in the HAR model to find the optimum set-up:

a) *Gyrometer data of all three sensors:* the modality of this data combination is nine.
b) *Accelerometer data of all three sensors:* the modality of this data combination is nine.
c) *Accelerometer and gyrometer data of all three sensors:* the modality of this data combination is 18.
d) *Ankle gyrometer and accelerometer data:* the modality of this data combination is six.
e) *Hand gyrometer and accelerometer data:* the modality of this data combination is six.
f) *Chest gyrometer and accelerometer data:* the modality of this data combination is six.
g) *Hand and ankle gyrometer data:* the modality of this data combination is six.
h) *Hand and ankle accelerometer data:* the modality of this data combination is six.
i) *Hand and ankle gyrometer and accelerometer data:* the modality of this data combination is 12.
j) *Chest and ankle gyrometer data:* the modality of this data combination is six.
k) *Chest and ankle accelerometer data:* the modality of this data combination is six.
l) *Chest and ankle gyrometer and accelerometer data:* the modality of this data combination is 12.
m) *Hand and chest gyrometer data:* the modality of this data combination is six.

TABLE 1. *8-fold cross-validation accuracy and its standard deviation of the proposed model on different signal combination.*

| Input Data | val_accuracy | val_acc_std |
|---|---|---|
| Chest and Ankle IMU (l) | 99.98 | 0.02 |
| Hand and Chest IMU (o) | 99.95 | 0.08 |
| Chest and Ankle Gyrometer (j) | 99.94 | 0.09 |
| Chest and Ankle Accelerometer (k) | 99.92 | 0.09 |
| Hand and Ankle IMU (i) | 99.89 | 0.15 |
| All 3 IMUs (c) | 99.77 | 0.42 |
| Gyrometer (a) | 84.49 | 2.84 |
| Accelerometer (b) | 83.32 | 4.28 |
| Hand and Chest Gyrometer (m) | 80.85 | 4.01 |
| Ankle IMU (d) | 80.77 | 6.92 |
| Hand and Chest Accelerometer (n) | 79.40 | 6.15 |
| Hand and Ankle Gyrometer (g) | 76.23 | 7.56 |
| Hand IMU (e) | 75.58 | 9.64 |
| Hand and Ankle Accelerometer (h) | 73.78 | 5.30 |
| Chest IMU (f) | 73.56 | 9.77 |

n) *Hand and chest accelerometer data:* the modality of this data combination is six.
o) *Hand and chest gyrometer and accelerometer data:* the modality of this data combination is 12.

The IMU sensors record linear and angular accelerations in three dimensions. Hence the modality of each sensor is three. For instance, the full IMU signals of the hand sensor have a modality of six.

The selected data has been mapped to a zero mean and standard deviation of one to minimise the effect of unit and amplitude and signal dominance on the model. Then, the multi-modal time series was transformed into 3D tensors for the deep learning model. The window size for this transformation was one second for the signals to be long enough in time to hold enough information related to the activity and short enough for real-time application. Each window has a 25% slide and 75% overlap with the previous window.

### III. MODEL DEVELOPMENT

In this study, we aim to find the best combination of inertial input signals located on a person's hand, ankle, and chest. We proposed a model based on Deep Convolutional Neural Networks (DCNN) to achieve the best performance. Convolutional layers are capable of extracting the local features of the signal. Stacking multiple convolutional layers helps the network to detect higher levels of features.

As shown in Fig 1, the proposed CNN model consists of two one-dimensional convolution layers followed by two hidden Fully Connected (FC) and an output layer. The first convolution layer has 16 filters with a kernel size of seven. The second one has 32 filters with a kernel size of 11. A max-pooling layer follows each convolution layer to reduce the network size. The hidden FC layers use 32 and 24 neurons, respectively. All layers except the output layers use Rectified Linear Unit (ReLU) activation function. The number of neurons in the output layer is equivalent to the number of activities and softmax activation function. A 0.3 dropout rate has been applied to all layers to prevent overfitting. The proposed model can have between 26565 and 27525 trainable parameters depending on the input signals' modality.

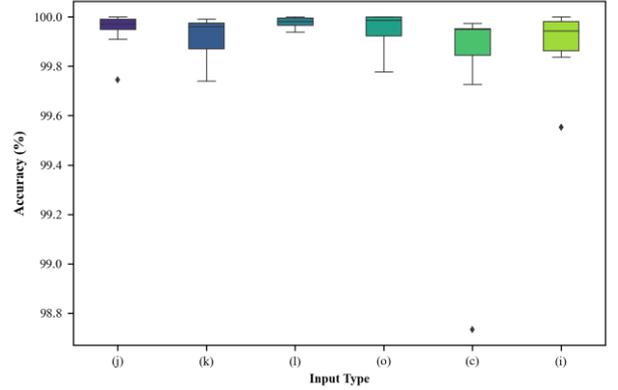

*Fig. 2. Cross-entropy accuracy distribution of the CNN model over the top perfoming data combinations.*

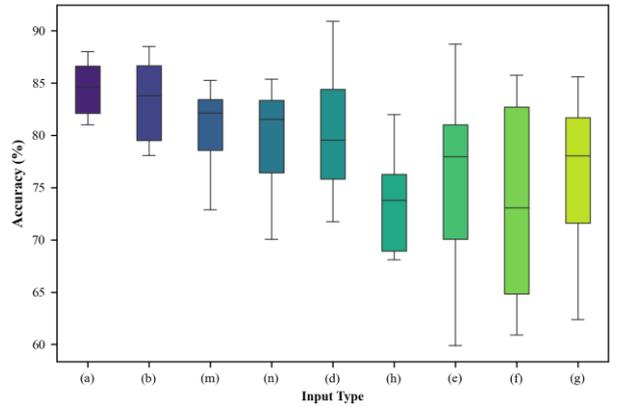

*Fig. 3. Cross-entropy accuracy distribution of the CNN model over other data combinations.*

The cross-entropy loss function and Adam optimiser have been used to train the models. The k-fold cross-validation method has been used for testing and training these models. The train and test data have been split based on subjects to block any data leakage in each step. The training will continue until there is no significant improvement in the validation loss value for an extended period of time. Next section discusses the performance results.

### IV. RESULTS AND DISCUSSION

The CNN model has been trained on all proposed combinations of inertial data in an 8-fold cross-validation approach. In each step, the validation data was one of the eight subjects. Table 1 exhibit the 8-fold cross-validation accuracy of the proposed CNN models. Model performance on the chest and ankle gyroscope and accelerometer data plus all other signal combinations containing more than one sensor's full IMU information were in a different league compared to the other signal combinations. The highest performance was achieved by using the full IMU data of the chest and ankle, with 99.98% cross-validation accuracy, followed by the hand and chest's full IMU signals at 99.95% accuracy. The lowest performance in this league is for all three sensors' full IMU data with 99.77% accuracy.

The mentioned inputs resulted in consistent performance in the proposed model, with all of them showing minimal deviation in their results. The standard deviation of cross-validation accuracy of the model on each data combination is shown on table 1.

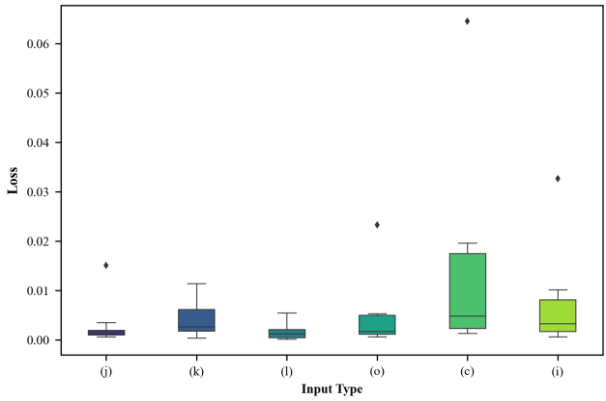

Fig. 4. Minimum cross-entropy loss of the proposed model on top performing signal combinations.

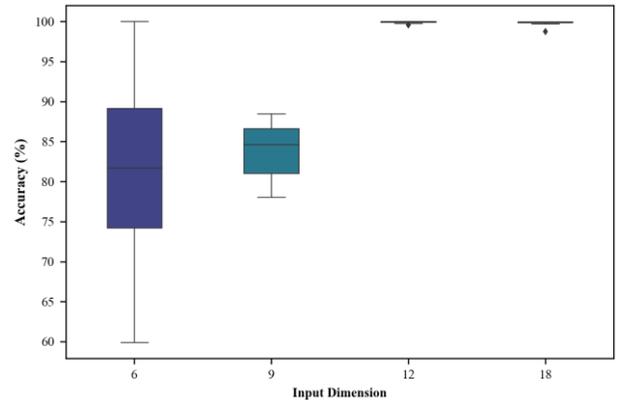

Fig. 6. The performance distribution of the proposed model trained on data with different modalities.

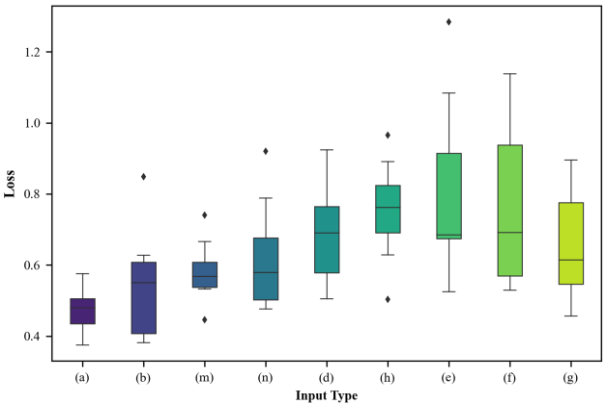

Fig. 5. Minimum cross-entropy loss of the proposed model on other signal combinations.

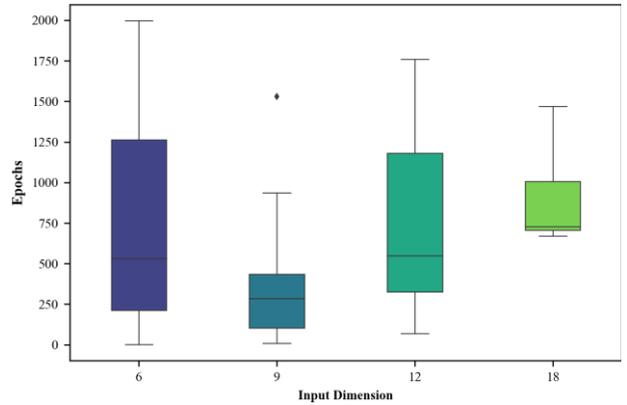

Fig. 7. The number of training epochs required for each signal combination.

Other combinations of data lead to a prediction accuracy of between 73 to 85%. In this lower performing group, angular and linear acceleration data of all three sensors showed better performance, demonstrating 84.5 and 83.3% cross-validation accuracy, respectively. Ankle full IMU signals had the highest model performance among single sensor data. Contrary, the chest sensor's signals lead to the lowest performance of all proposed signal combinations.

The increasing trend in the performance with the dimensionality of the data is another interesting observation. While input data with the modality of six performed with 82.2% accuracy, nine-modal signals lead to 83.9% accuracy. This number increases to 99.94% for 9-modal time-series signals. However, we did not observe any improvement in the performance by further increasing the signal's modality. One possible explanation for this phenomenon is that perhaps most signal combinations with the modality of 12, retain enough information for the deep learning model to detect the activity, and adding more signal channels (more modality) will not add much to the data information. In contrast, this increase in dimensionality adds to the complexity of the final model. Not only does the increased number of trainable parameters in the model require more data for training, but also it adds to the complexity and computation cost in the prediction stage.

This increase in the model complexity also shows itself in the training epochs. Other than signal combinations with the modality of six, with the increase in the dimensionality of the input, the model needs more training loops to optimise its weights. While, on average, it takes 378 epochs to train the model on the data with the modality of nine, this number increases to 722 and 878 for the inputs with the modality of 12 and 18, respectively. However, input signals with the modality of six showed a variety of different and inconsistent results, with the average number of training epochs being 744 and a standard deviation of 612. This can be due to the fact that most signal combinations with the modality of six do not contain enough information for the deep learning model to optimise its weights quickly. However, hand and chest sensors' angular and linear acceleration data are exceptions.

## V. CONCLUSION

This study compared the effect of input signal combination and sensor location on HAR performance. A deep convolutional neural networks model has been proposed and trained on a different combination of linear and angular acceleration signals of a person's dominant hand, leg and chest. Five slow-paced mundane activities have been selected to help control or adjust the wearable assistive device; sitting, standing, walking, ascending and descending stairs.

The proposed CNN model was trained on random one-second portions of different signal combinations. One-leave-out k-fold cross-validation method has been used to evaluate the performance. The test and train data, in each step, has been selected based on participants to stop data leakage. The chest and ankle full IMU data demonstrated the best performance in the model, followed by the chest and hand's full IMU signals.

The worst results belong to the chest's full IMU signals and hand and ankle accelerometer signals. Overall, all signal combinations with the modality of 12 and more performed with exceptional accuracy. Other data combinations with a modality of nine or less had shown an accuracy of less than 85%, except angular and linear acceleration data of hand and chest sensors.

For future works and research directions, one can implement smart algorithms to reduce the dimensionality of the proposed data combinations without losing much information.